\documentclass[prb,preprintnumbers,amsmath,amssymb,floatfix,aps,twocolumn]{revtex4}
\usepackage{graphicx}

\usepackage{xcolor}

\usepackage{amsmath}
\usepackage{amsbsy}
\usepackage{gensymb}
\usepackage{float}
\usepackage[utf8x]{inputenc}

\begin{document}

\title{Strain-controlled thermoelectric properties of phosphorene 
hetero-bilayers}

\date{\today}

\author{J. W. Gonz\'alez}
\email{jhon.gonzalez@usm.cl}
\affiliation{Departamento de F\'{i}sica, Universidad 
T\'{e}cnica Federico Santa Mar\'{i}a, Casilla Postal 
110V, Valpara\'{i}so, Chile.}

\begin{abstract}
The application of strain to 2D materials allows manipulating the electronic, magnetic, and thermoelectric properties. These physical properties are sensitive to slight variations induced by tensile and compressive strain and to the uniaxial strain direction. 
Herein, we take advantage of the reversible semiconductor-metal transition observed in certain monolayers to propose a hetero-bilayer device.  We propose to pill up phosphorene (layered black phosphorus) and carbon monosulfide monolayers. In the first, such transition appears for positive strain, while the second appears for negative strain.
Our first-principle calculations show that depending on the direction of the applied uniaxial strain; it is possible to achieve reversible control in the layer that behaves as an electronic conductor while the other layer remains as a thermal conductor.  The described strain-controlled selectivity could be used in the design of novel devices.
\end{abstract}

\maketitle

%%%  electron or hole doping 
\section{\label{sec:intro} Introduction}
With the recent advances in the research of two-dimensional materials, a plethora of exciting new phenomena have been predicted and measured\cite{dai2019strain,zhu2017recent}. 
Graphene was the first of a long list of synthesized and predicted 2D-materials that could impact future technologies.
Among others, their electronic properties and  control in layered materials have been 
explored, including few-layers compounds from the carbon (group 14), nitrogen (group 15), 
and oxygen (group 16) groups\cite{shah2020experimental,zhu2020kagome,guo2017thermoelectric}.
A significant challenge in designing practical devices based on low-dimensional materials 
is to archive precise and systematic control of the band gap.  
In particular, some promising solutions have been proposed in few-layers systems; the combination 
of layers of different materials\cite{ajayan2016van,nutting2020heterostructures}, 
the use of stain\cite{zhu2016black,dai2019strain,leon2020strain,aslan2018probing}, the stacking 
control\cite{dai2014bilayer,gao2012artificially}, or the introduction of intercalated 
impurities\cite{lin2018research,cortes2018stacking}.

\begin{figure}[!hb]
\centering
\includegraphics[clip,width=1\columnwidth,angle=0]{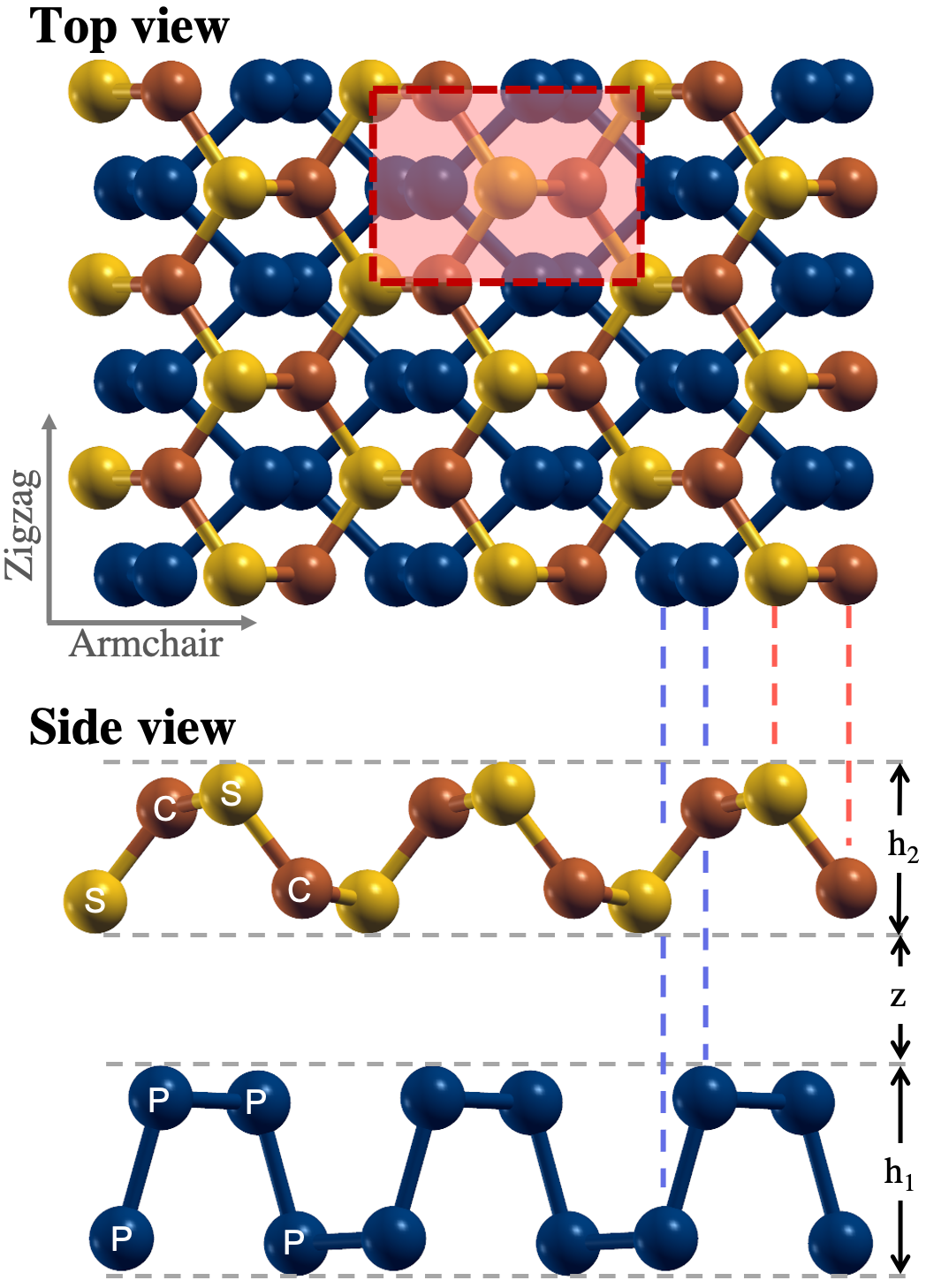} 
\caption{(Color on-line) 
Ball-and-stick  representation  of the BP-CS hetero-bilayer with top and side views. 
The bottom phosphorene layer (BP-layer) is composed of phosphorous atoms represented with blue spheres. The top carbon monosulfide layer (CS-layer) includes carbon atoms as brown spheres and sulfur atoms as yellow spheres. The carbon atoms are located in the inner part of the CS-layer. The layer-layer separation $z$ is $3.1$ \AA{} and the lattice constants are $a = 4.25$ \AA{} and $b = 3.04$ \AA{}.
The unit cell with eight atoms is highlighted with a dashed line.
% We oversize the lines to indicate the distances to ease readability.
}
\label{Fig:scheme}
\end{figure}

In the few-layer systems, the strain can be applied indirectly by thermal or mechanical manipulation 
of the substrate or directly by mechanical deformation\cite{dai2019strain,frisenda2017biaxial,caneva2018mechanically,zhu2016black}.
%%%%
Previous studies show that most 2D materials can easily overcome strain values above $\pm10$ \% without breaking\cite{bertolazzi2011stretching,caneva2018mechanically,frisenda2017biaxial}; for instance, graphene can tolerate values above 25 \% and recover its original structure\cite{lee2008measurement}.
The response to applying strain to 2D materials depends on both the crystallographic structure and 
the composition of the material. For instance, although vanadium nitride (V$_2$N) and vanadium 
carbide (V$_2$C) have a similar crystal structure, compressive strain applied to V$_2$N monolayers increases 
the magnetic moments, while the same effect can be achieved using tensile strain on V$_2$C 
monolayers\cite{gao2016monolayer}.
In some monolayers like the molybdenite (MoS$_2$) and carbon monosulfide (CS), the band gap decreases 
with the tensile strain\cite{gonzalez2019highly,scalise2012strain}.
However, the tensile strain increases 
the band gap in monolayers of graphene, phosphorene, and group-IV 
monochalcogenides\cite{si2016strain,lv2014enhanced,rodin2014strain}.

Herein, we study the electronic and thermoelectric properties of a system made by two monolayers with similar structures but opposite band gap response upon the applied strain.  
In this work, we study the electronic and thermoelectric properties 
using the density functional theory (DFT) calculations.
We consider a Van der Waals hetero-bilayer consisting of a carbon monosulfide monolayer (CS-layer) with a phosphorene monolayer (BP-layer), and we focus on the thermoelectric response upon the strain. 
Both monolayers are semiconductors\cite{li2017direct,lv2014enhanced,alonso2017stable,gonzalez2019highly}, the phosphorene 
has a direct band gap of $2.0$ eV, and the carbon monosulfide monolayer presents an indirect band gap of $1.1$ eV. 
On one side, we have a single layer of black phosphorus, also called phosphorene 
(BP-layer) --bottom layer, fig. \ref{Fig:scheme}--, and on the other side, we have a monolayer of carbon monosulfide (CS-layer) --top layer, fig.  \ref{Fig:scheme}--. 
Both layers present a black-phosphorus-like atomic structure and because of its two-dimensional puckered honeycomb structure, where each atom is bound to three neighbors, 
we anticipate a composite material with high mechanical flexibility and highly anisotropic electronic 
properties\cite{zhu2016black,jain2015strongly,gonzalez2019highly,lv2014enhanced}.

\section{Metodology }
Within the density functional theory framework, we employ the plane-wave self-consistent approach as 
implemented in the  {\sc Quantum ESPRESSO} package\cite{giannozzi2009quantum}. 
We use a generalized gradient approximation in the Perdew-Burke-Ernzerhof exchange-correlation (GGA-PBE) 
functionals\cite{perdew1996generalized,pseudo}. 
We include the long-range dispersive forces with the DFT vdW-D3 approach to treat Van der Waals 
interactions\cite{grimme2010consistent,goerigk2017comprehensive}. 
Self-consistent charge calculations are converged up to a tolerance of $10^{-8}$. To avoid self-interaction 
in the out-of-plane direction, we add 20~\AA{} of empty space in the z-direction.  
To sample the Brillouin zone, we use a dense k-grid of $20 \times 20 \times 1$ Monkhorst-Pack, and the 
wave functions are expanded in plane waves until a kinetic energy cutoff of $680$ eV. 
The atomic positions are allowed to relax within the conjugate gradient method until forces are converged 
with a tolerance of $10^{-3}$ eV/\AA{}. 

%% GW and similars
We calculate the thermoelectric coefficients using the PBE-DFT band structure, 
which can be used as a lower band gap limit\cite{fei2014enhanced}.
Although an accurate theoretical description of band gap requires a 
semi-local approach as the GW 
approach\cite{tran2009accurate,tran2014layer}. 
Under experimental conditions, the electronic properties 
are affected by several external factors that modify the 
band gap, becoming smaller than the calculated with 
sophisticated theoretical approaches\cite{hybertsen1986electron}.

The thermoelectric coefficients are calculated within the semi-classical Boltzmann 
transport theory within the constant relaxation time approximation (RTA), as implemented in the 
{\sc BoltzTraP} code\cite{madsen2006boltztrap}. 
In this approximation, the relaxation time is constant ($\tau_0 = 1$ fs) and therefore the Seebeck coefficient is independent of it.
This approximation has been successfully used to describe the transport coefficients 
of a wide range of thermoelectric materials\cite{gonzalez2019highly,li2016anisotropic,hung2015diameter}. 
For the calculation of the transport coefficients a denser k-grid of 
$50 \times 50 \times 1$ Monkhorst-Pack is used to sample the Brillouin zone.

The thermoelectric coefficients directly depend on the temperature $T$ and 
chemical potential $\mu$. The Seebeck coefficient reads 
as\cite{madsen2006boltztrap,scheidemantel2003transport}
\begin{equation}
    S=\frac{q k_B}{\sigma} \int d\varepsilon 
    \left( -\frac{\partial f}{\partial \varepsilon} \right)
    \Xi \left( \varepsilon\right) 
    \frac{ \varepsilon - \mu }{k_B T},\label{eq:S}
\end{equation}
where $\sigma$ represents the electronic conductivity, $\sigma$ can be expressed as
\begin{equation}
    \sigma = q^2
    \int d\varepsilon 
    \left( -\frac{\partial f}{\partial \varepsilon} \right)
    \Xi \left( \varepsilon\right).
\end{equation}
The transport distribution, $\Xi$, can be written as
\begin{equation}
    \Xi = \sum _{\vec{k}} \vec{v}_{\vec{k}}\vec{v}_{\vec{k}} 
    \tau_{\vec{k}},\label{eq:SS}
\end{equation}
where $f$ is the Fermi-Dirac distribution, $q$ is the charge of 
the carriers, 
$\tau_{\vec{k}}=\tau $ is the relaxation time
and $\vec{v}_{\vec{k}} = \frac{1}{\hbar} \left( \partial \varepsilon_{\vec{k}} / \partial \vec{k} \right)$
is the group velocity of the $\vec{k}$ state.

The most significant changes in the thermal coefficients (Eqs. \ref{eq:S} to \ref{eq:SS}) could be associated with the band gap strain-tunability and modifications in the Fermi velocity (i.e., band curvature) of the subbands around the charge neutrality point.

\section{Results and Discussion}

\subsubsection{Equilibrium configuration}

We explore the four possible highly symmetric configurations\cite{zhang2015stacked}. In AA stacking, the top layer is vertically displaced over the bottom layer (spatial group Pmna). In AB stacking, there is a relative displacement of half a lattice vector along the zigzag (or armchair) direction (spatial group Pbcm). The AA' stacking is defined by the relative displacement in one lattice vector along the zigzag (or armchair) direction; thus, the layers are mirror images of each other (spatial group Pmma). Finally, the AB' stacking is characterized by a relative displacement of 3/2 lattice vector along the zigzag (or armchair) direction (spatial group Pccm).
We find that the most stable configuration is the AB staking, being the same staking of the bulk and few-layers black-phosphorus \cite{shu2016stacking,zhang2015stacked}.
We summarize our main results in table \ref{table_stacks}.

\begin{figure}[!ht]
\centering
\includegraphics[clip,width=\columnwidth,angle=0]{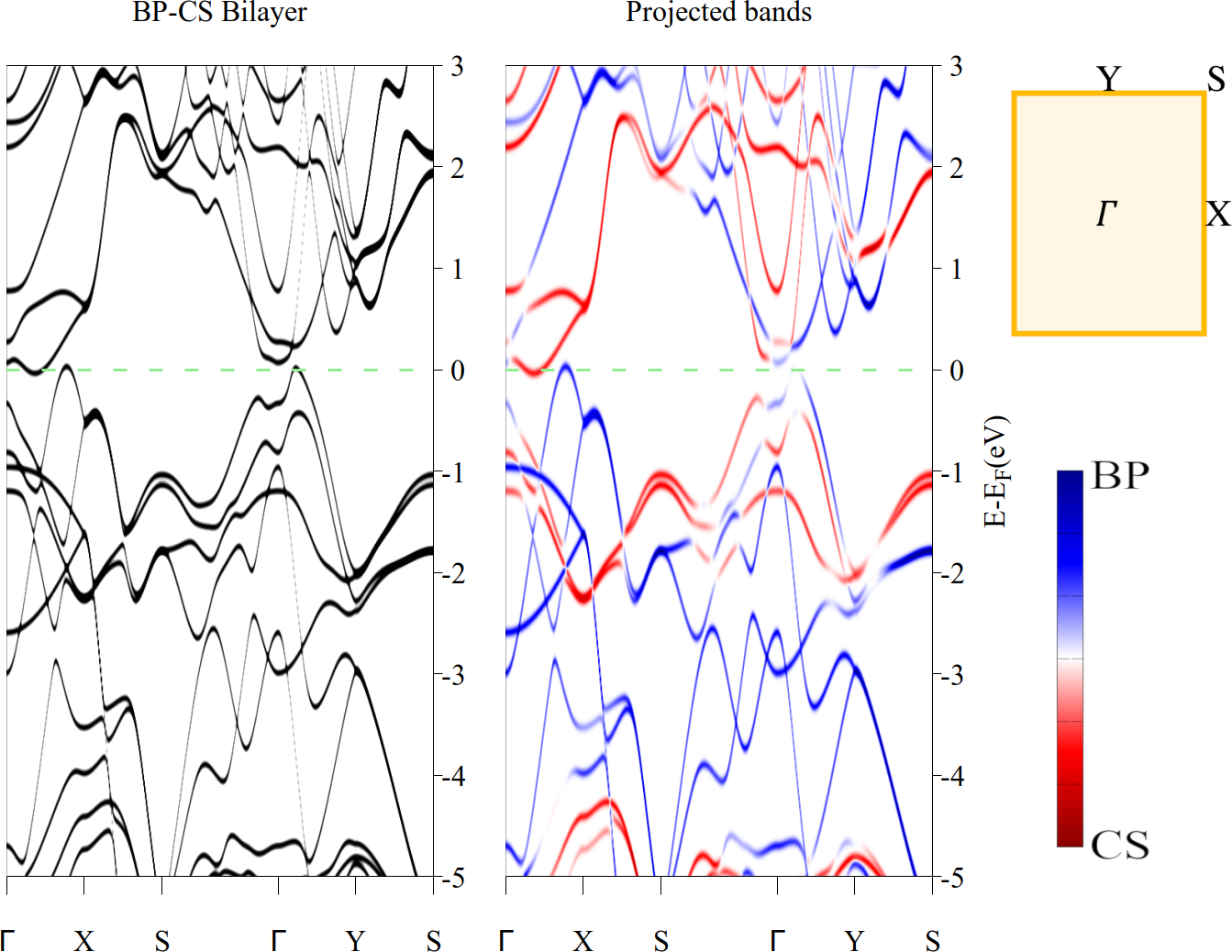} 
\caption{(Color on-line) Band structure of the BP-CS bilayer in the left panel. 
In the right panel,  the layer-projected band structure represents the dominant layer contribution.  The blue subbands have a dominant BP-layer contribution, and the red subbands have a dominant contribution from the CS-layer.
Inset shows the high-symmetry points in the reciprocal space. 
%The lengths of $\Gamma-X$ and $\Gamma-Y$ paths are the same when the lattice vectors $a$ and $b$ are equal.
}
\label{Fig:bands_0}
\end{figure}

\begin{center}
\begin{table}[h!]
\begin{tabular}{|c|c|c|c|c|}
\hline 
 & $\Delta E $ (meV) & $z$ (\AA{}) & $a$ (\AA{})& $b$ (\AA{})\\ 
\hline 
 AA  & 64.6 & 3.36 & 4.196	& 3.034\\ 
\hline 
 AA' & 84.8 & 3.48 & 4.203	& 3.035\\ 
\hline 
 AB  & 43.0 & 3.24 & 4.224	& 3.034\\ 
\hline 
 AB' & 0.0  & 3.07 & 4.254	& 3.034\\ 
\hline 
\end{tabular} 
\caption{Stability order and geometric parameters of the different BP-CS bilayer stackings. The energy difference relative to the ground state (AB stacking), the interlayer separation $z$ (defined in fig. \ref{Fig:scheme}), and the magnitude of lattice vectors a, b.}
\label{table_stacks}
\end{table}
\end{center}

The AB-like stakings have the lowest energy (table \ref{table_stacks}); these configurations favor compact structures. This behavior has also been observed in other 2D 
systems\cite{cortes2018stacking,leon2020strain,gonzalez2010electronic}.
The variations in the lattice vectors in table \ref{table_stacks} are small enough to be negligible, thus have no impact on the stability order of the system.
The interlayer distance and the lattice vectors for the different BP-SC stackings follow the same trend as those shown by their equivalent in phosphorene bilayers\cite{zhang2015stacked}.

% MD
For the non-strained system, the phonon dispersion calculation has no negative frequencies. Moreover, as a double-check of  structure stability, we perform a Born-Oppenheimer molecular dynamics calculations\cite{dai2014structure,yang2015new} to prove that the structure at 0\% strain is stable at room temperature ($300$ K) after $2.5$ ps of thermalization.

For the isolated phosphorene monolayer (BP-layer) we find a lattice constant value 
$a = 4.58 $ \AA{} and $b = 3.30 $ \AA{}, for an aspect 
ratio $a/b = 1.39$. In the CS monolayer we find a value $a = 4.01 $ \AA{} and $b = 2.78 $ \AA{}, for an 
aspect ratio  $a/b = 1.44$. Our lattice constants and geometries are similar to those reported in 
previous works\cite{castellanos2014isolation,shu2016stacking,alonso2017stable,gonzalez2019highly}.
%mix
The lattice vectors in the equilibrium configuration of our BP-CS bilayer 
deviate slightly from the average values between the two monolayers, being  
$a = 4.245 $ \AA{} and $b= 3.034$ \AA{} for an aspect ratio $a/b = 1.40$. 
As a reference, the BP-layer in the hetero-bilayer is biaxially compressed in about $-8\%$, 
keeping the original phosphorene aspect ratio.
%
% Alturas y geometria
Using the geometrical parameters shown in fig. \ref{Fig:scheme}, the ground-state 
geometry is characterized by a layer-layer separation of $z= 3.14$ \AA{}, a layer 
thickness of $h_1= 2.28$ \AA{} and $h_2= 1.69$ \AA{}, 
and we also find a slight buckling variation of $dh_1= 0.08$ \AA{} for BP-layer and 
$dh_2= 0.21$ \AA{} for the CS-layer. 
Note that in the latest, the sulfur atoms are positioned in the outer part of the layer.

% Bandas
In fig. \ref{Fig:bands_0} we show the band structure of the BP-SC bilayer (left panel) and its projection on atomic orbitals of each layer (right panel). For every energy and k-point, we evaluate the difference between the sum of projections on atomic orbitals of each layer.
Without strain, the BP-CS bilayer is semimetal. In fig. \ref{Fig:bands_0}, we observe
crossovers of the conduction and valence bands along the $\Gamma$-$Y$ and $\Gamma$-$X$ points. 
The analysis of the projected layered band structure in the right panel of fig. \ref{Fig:bands_0} reveals that the crossovers have different characters. On the one hand, the band crossover between the $\Gamma$-$Y$  points involves both layers. On the other hand, the band crossover observed between $\Gamma$-$X$ points is characterized by a BP-dominant valence band and a CS-dominant conduction band. This separation allows possible carrier-layer separation effects. 
Around the Fermi level, the last valence band and first conduction bands are localized on different layers. Thus the lowest energy electron-holes pairs are spatially separated, making recombination processes even more unfavored in energy\cite{kosmider2013electronic}.

\begin{figure}[t]
\centering %rkpdos_0_decoupled2.png
\includegraphics[clip,width=0.83\columnwidth,angle=0]{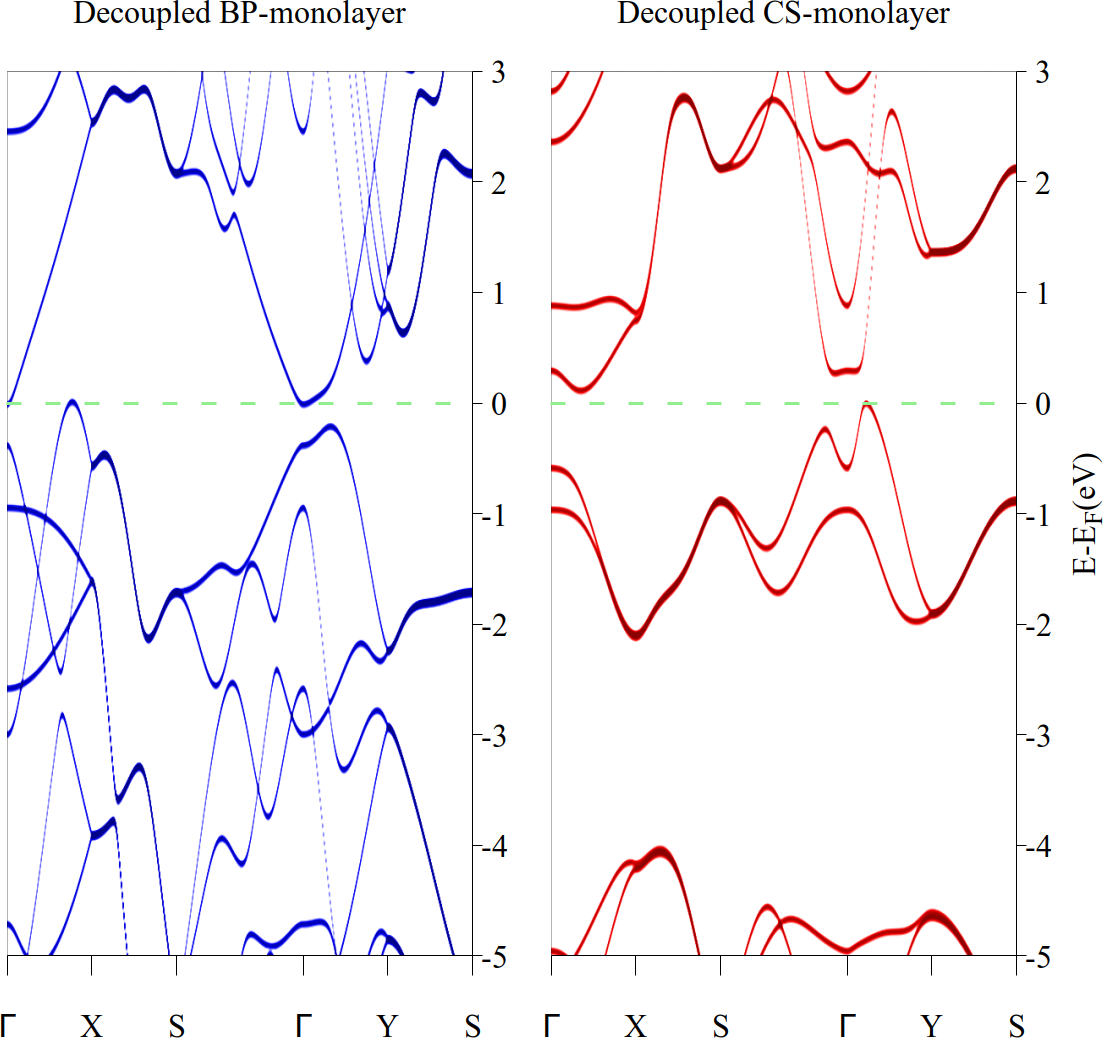} 
\caption{(Color on-line) Band structure in the decoupled limit. 
In the decoupled limit, we use the same lattice vectors and atomic positions of the BP-CS bilayer but removing one layer.
In the left panel, the band structure of the decoupled BP-layer and in the right panel, the band structure of the decoupled CS-layer. }
\label{Fig:bands_0_LL}
\end{figure}

% decoupled limit
To confirm and trace the features shown by the projected band structure in layers, we calculate the band structure of the system in the decoupled limit (fig. \ref{Fig:bands_0_LL}). In this limit, we fix the crystal parameters and remove one of the layers.
The projection of the band structure by layers and the analysis of the band structure in the decoupled limit confirm the existence of zones where the states are markedly localized in one layer or the other.
The effects of layer-layer interactions prevent a direct one-to-one comparison of the band structure of the bilayer with its separate components.
Although, the BP-CS bilayer system is semimetal (fig. \ref{Fig:bands_0}). 
In the limit of the decoupled layers, the BP-layer is  semimetal 
(fig. \ref{Fig:bands_0_LL} right), whereas the CS-layer is  semiconductor with a narrow indirect band gap  (fig. \ref{Fig:bands_0_LL} left).

\begin{figure}[!ht]
\centering
\includegraphics[clip,width=1\columnwidth,angle=0]{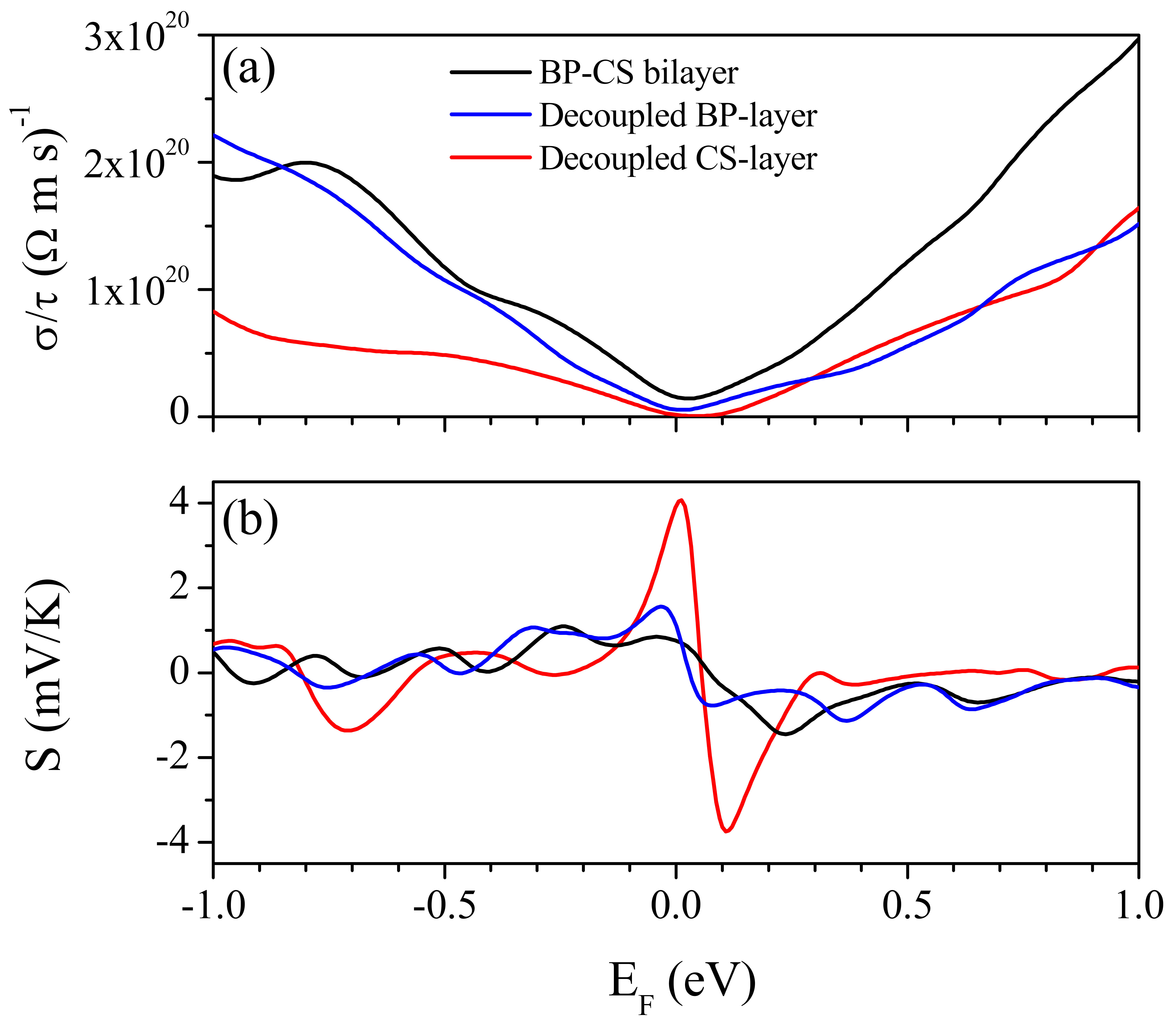} 
\caption{(Color online)  Thermoelectric properties at room temperature 
($300$ K) of BP-CS bilayer (black line), decoupled BP-layer (red line), 
and decoupled CS-layer (blue line) without strain. 
The electronic conductivity ($\sigma/\tau$) in top panel, and the Seebeck 
coefficient (S) in bottom panel.}
\label{Fig:thermo_0}
\end{figure}

%Propiedades termoelectricas
The comparison of the thermoelectric properties in the bilayer 
with its components at the limit of decoupled layers, reveals 
that the controlled manipulation of layer-layer separation (in z axis) becomes an 
efficient way to tune-in the electronic and thermoelectric properties. 
The effect can be observed in electronic conductivity ($\sigma$) and 
Seebeck coefficient ($S$, also known as thermopower) physical properties shown in 
fig. \ref{Fig:thermo_0}. The transition between bilayer and decoupled limit 
is smooth and continuous\cite{cortes2018stacking}. 
For instance, the electronic conductivity in fig. \ref{Fig:thermo_0}(a) 
shows the bilayer as a good conductor in an energy range around the 
Fermi level\footnote{We consider a $\pm0.5$ eV window around the Fermi level.}, while its components are poor conductors ($\sigma \rightarrow 0$). Note that the conductivity in the bilayer is higher than the sum of its components.
In contrast to the tendency observed in the conductivity in 
fig. \ref{Fig:thermo_0}(b), the sum of two good thermoelectric conductors 
produces a poor thermoelectric conductor. 
The decoupled CS-layer presents a high Sebeek coefficient, with a maximum 
around the Fermi level of $4.1$ mV/K, and the BP-layer reaches the 
$1.6$ mV/K. In contrast, at the Fermi level, the Sebeek coefficient of 
the BP-CS bilayer $1.4$ mV/K.

\subsubsection{Strain-induced structural phase transition}
% ENERGIA STRAIN
We define the tensile strain/compression as the percentual change in the lattice vectors relative to the non-strained BP-CS unit cell. The crystal cell is stretched with positive values, and the cell is compressed for negative strain values. 
We consider changes in the unit cell in the armchair (along the x-axis) and zigzag (along the y-axis) directions. 
For each strain value, we start with the non-strained atomic positions in crystal coordinates and let the atomic position relax.

Analyzing the behavior of the total energy of the BP-CS bilayer upon strain, we 
identify some discontinuities (figure not included). Interestingly, for a 
$\gtrsim +10\%$ strain in the zigzag direction, we observe an abrupt change in 
the monotony of the energy curve as a function of the 
strain; we can identify the same behavior when the strain is biaxially applied. 
We associate such changes with a phase change in the CS layer. This phase is metastable, 
and it corresponds to the $\kappa$ phase of the CS-layer\cite{alonso2017stable}. 
For strain values of $\lesssim-10\%$, we also observe a structural phase change, but this time in the 
BP-layer. The change in the BP-layer structure at $\lesssim-10\%$ strain yields to a crystal 
in a double-decker hexagonal structure, similar to K4-phosphorus\cite{liu2016phosphorus}.
We also check the appearance of both phase-transitions in a $2\times2$ supercell, but 
further structure stability analyses beyond $\pm 10\%$ strain are out of the scope of 
this manuscript.

\begin{figure*}[!ht]
\centering
\includegraphics[clip,width=.9\textwidth,angle=0]{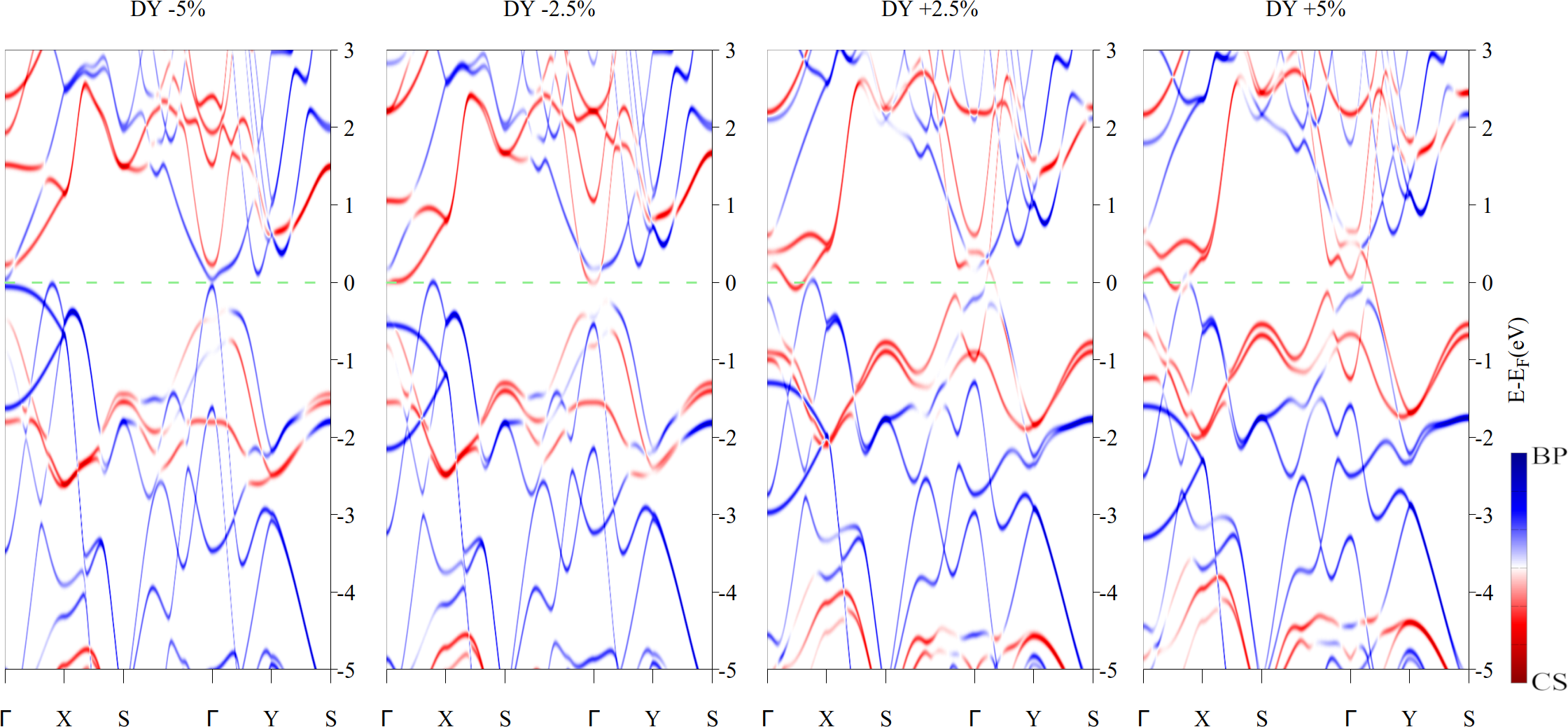}
\caption{(Color online) Projected band structure of the BP-CS bilayer for several monoaxial strain cases in the zigzag direction (along the y-axis). Positive values correspond to a system stretching, and negative values correspond to a system compression.
For positive stain values, note the metallic character induced on the CS-layer, revealed as a gap-closing along the $\Gamma-Y$ path.
For negative strain values, we observe a change in the electronic state to a semi-metal phase. %% En los limites tenemos una capa metalica o la otra
}
\label{Fig:DY}
\end{figure*}

\subsubsection{Layer selectivity: strain in zigzag direction}
We observe several changes in the band structure that occur even for small strain values. 
Both the unidirectional and bidirectional strain can modify the band structure and, consequently, affect the thermoelectric properties. Due to the selectivity introduced when considering unidirectional tension in the zigzag direction, we first focus on this case.

%simetria
If we take as origin in the center of the unit cell (fig. \ref{Fig:scheme}), the crystal presents reflection symmetry around the xz-plane and is not present along the yz-plane. 
We associate sensibility to the strain in the zigzag direction, 
which dramatically affects the system's properties to this crystal symmetry. 
For example, when we apply an anisotropic tension along the zigzag direction, depending on the sign of the applied strain, we can manipulate the system so that one of 
the layers becomes metallic and, therefore, an excellent electronic conductor. At the same 
time, the other layer becomes a semiconductor and therefore presents a higher thermopower, meaning higher conversion of thermal into electrical energy capabilities.

When we apply a positive strain in the zigzag direction (fig. \ref{Fig:DY} right panels), the band structure of $+2.5 \%$ and $+5.0 \%$ reveal metallic systems. When looking at the band projection, we can identify the contribution of the BP-layer as a semiconductor (blue lines in fig. \ref{Fig:DY} right panels). At the same time, the CS-layer has a metallic character with bands crossing the Fermi level (red lines in fig. \ref{Fig:DY} right).
For negative strain in zigzag direction (fig. \ref{Fig:DY} left panels), the electronic band structure reveals a semimetallic character. 
The band projection on the CS-layer reveals a semiconductor with a gap of around $0.6$ eV, while the BP-layer presents a metallic character. In this case, the electronic transport can be achieved through the BP-layer, while the CS-layer became a thermal conductor.

When applying strain in the zig-zag direction (y-axis), the band structure changes between metallic and semi-metallic. An analysis of layer contribution reveals that the strain sign determines the narrowing of the energy gap of the bands belonging to one layer and the opening of the gap for the bands belonging to the other. 
From left to right in fig. \ref{Fig:DY}, the effect of strain in the zigzag direction can be separate in two parts. First, the energy gap associated with the BP-layer is $0.0$ eV for a strain of $-5.0$ \% and increases to $0.7$ eV at $+5.0$ \%. Second, the energy gap originated in the CS-layer is around to $0.6$ eV for $-5.0$ \% strain and becomes 0 for positive strain values.

\begin{figure}[!ht]
\centering
\includegraphics[clip,width=1.0\columnwidth,angle=0]{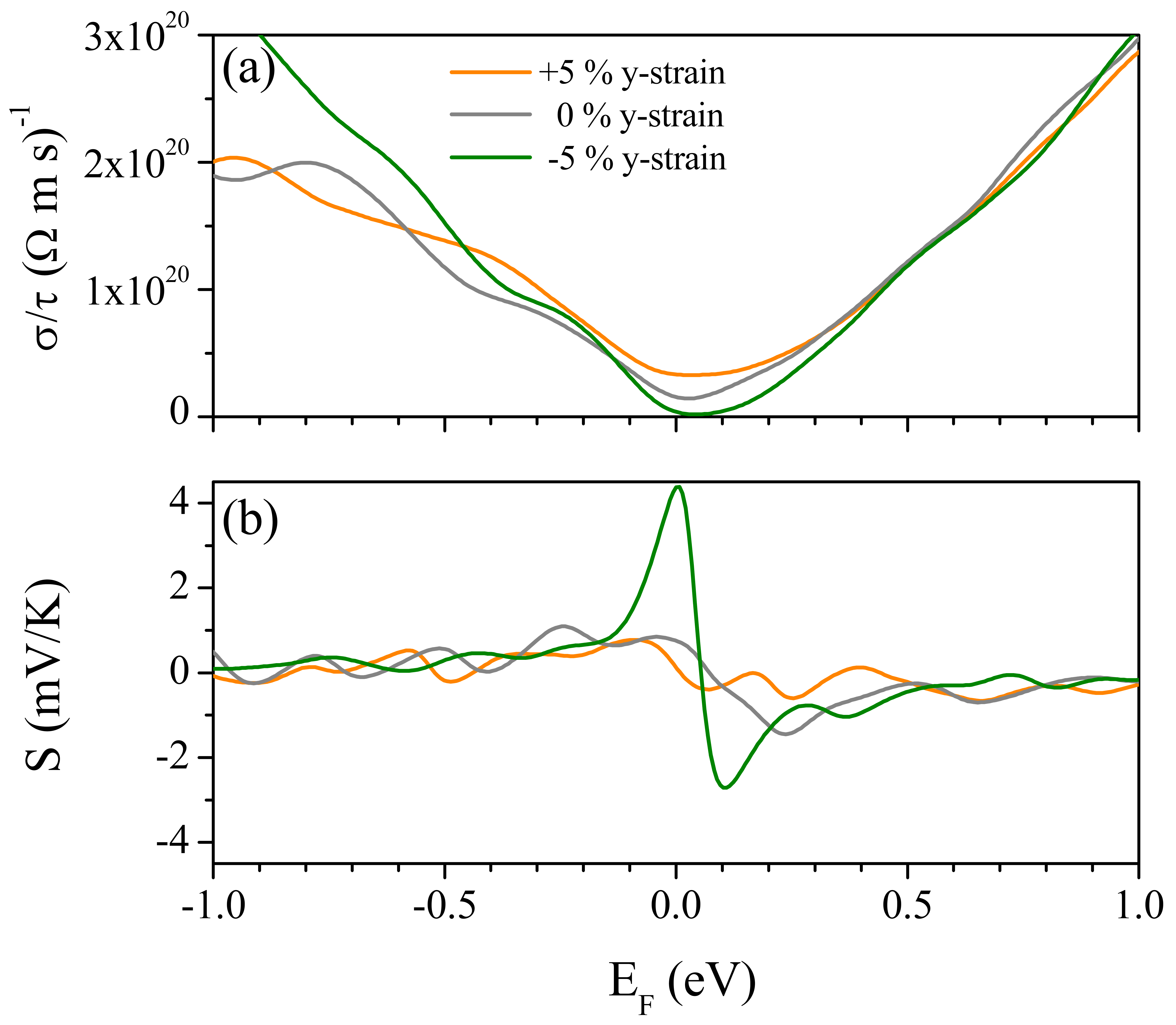}
\caption{(Color online) Thermoelectric properties at room temperature 
($300$ K) for the non-stained BP-CS bilayer (gray line), 
+5 \% stain in zigzag direction (orange line), 
and -5 \% stain in zigzag direction  (green line). 
The electronic conductivity ($\sigma/\tau$) in upper panel, and the Seebeck 
coefficient (S) in bottom panel.}
\label{Fig:termo_DY}
\end{figure}

Modifications observed in the band structure can be associated with 
the thermoelectric properties of the system\cite{gonzalez2019highly}.
In fig. \ref{Fig:termo_DY}, we compare the two main thermoelectric 
quantities, considering a strain-induced modification of $\pm 5\%$ 
in the zigzag direction.  For comparison, we have also included the 
equilibrium case (0\% strain). The manipulation via  applied 
strain can selectively make one of the layers metallic, improving 
(or worsening) each layer's electronic conduction capabilities.

\begin{figure*}[!ht]
\centering
\includegraphics[clip,width=0.9\textwidth,angle=0]{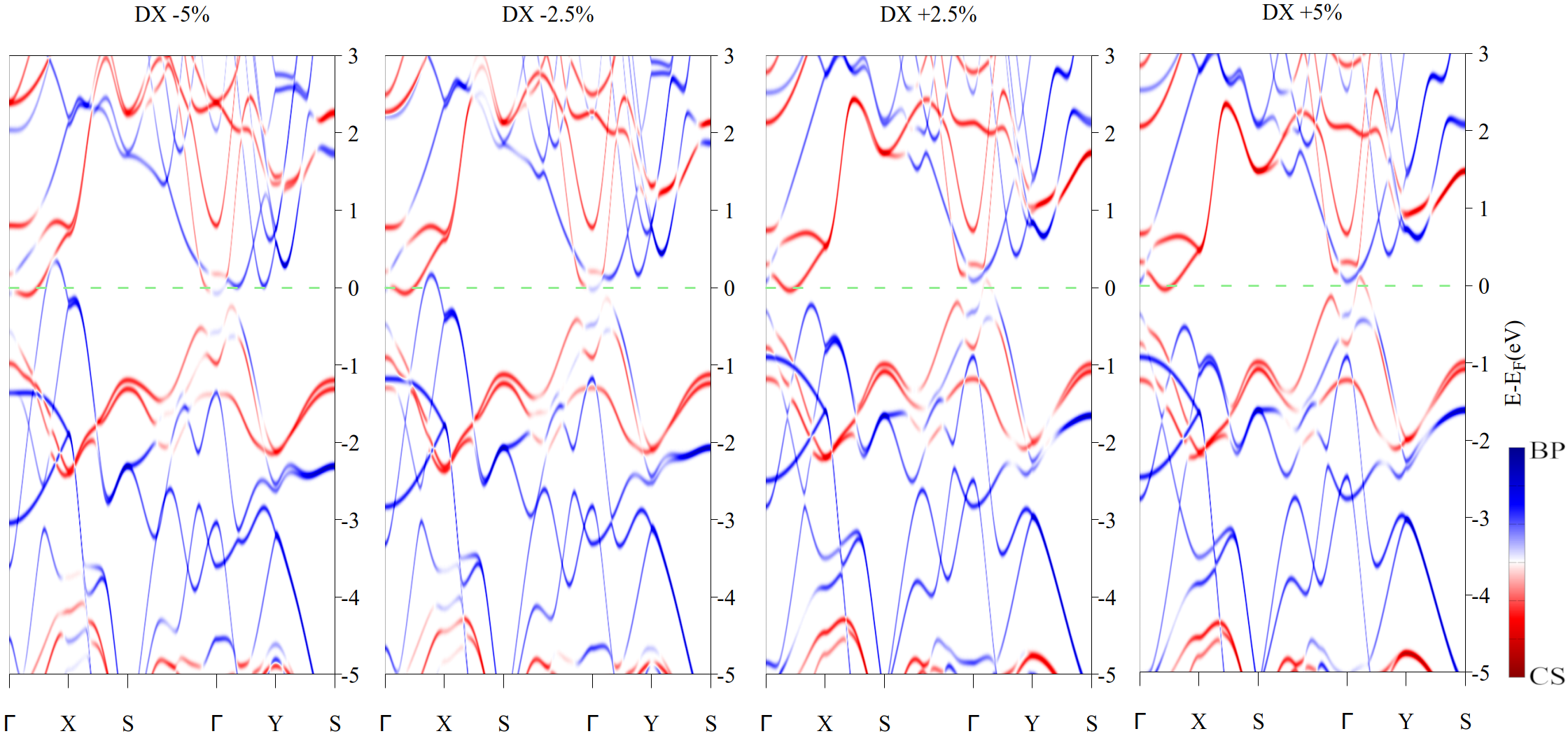}
\caption{(Color online) Projected band structure of the BP-CS bilayer
considering strain in the armchair-direction (along the x-axis). Figure distribution and colors as in fig. \ref{Fig:DY}.
}
\label{Fig:bnds_DX}
\end{figure*}

Given the strain-induced layer selective response, the bilayer electronic 
conductivity will depend on the layer that is currently acting as 
an electronic conductor.  In fig. \ref{Fig:termo_DY} (a), we  
compare the electronic conductivity response ($\sigma/\tau$) to strain. 
For $0$ \% strain, the minimum electronic conductance reaches $0.7  \times 10^{19} \, \left(\Omega m s\right)^{-1}$ and for a $+5$\% zigzag strain we obtain a minimum of $1.6  \times 10^{19} \, \left(\Omega m s\right)^{-1}$.
A band structure analysis (fig. (\ref{Fig:bnds_DX})) for 0\% and $+5$\% strain reveals that the electronic conductivity around the Fermi level is mainly performed through the CS-layer because of its strain-induced metallic character.

This behavior can associated 
with the metallic character of the CS-layer in for 0\% strain 
(fig. \ref{Fig:bands_0}) and in  $+5$\% zigzag strain (fig. \ref{Fig:DY}). 
The $-5\%$ zigzag strain conductivity curve decreases, showing a minimal conductivity tending to zero ($9.5 \times 10^{17} \, \left(\Omega m s\right)^{-1}$); under this strain, the BP-layer becomes metallic, and the CS-layer presents a band gap (fig. \ref{Fig:DY}). Since the conductivity is larger when the CS-layer presents a metallic behavior, we associate the decrease in the conductivity to the poorest 
electronic conductivity capabilities of the BP-layer.

The Seebeck coefficient (also known as thermopower) defined in equation \ref{eq:S} is affected by the size of the band gap and band curvature, both of which are strain-sensitive, especially near the charge neutrality point.
In fig. \ref{Fig:termo_DY} (b), we present the variation of the Seebeck coefficient (S) as a function of the chemical potential for several strain values in the zigzag y-direction.
The case with strain $-5\%$ in the zigzag direction presents a higher value than the other two cases in fig. \ref{Fig:termo_DY} (b), reaching a value of $\vert S_{max} \vert =4.4$ mV/K, value which is close to the $4.1$ mV/K found for the decoupled CS-layer. This larger Seebeck coefficient results from a larger band gap induced in the CS-layer by the compression of the system.
Depending on the electron or hole doping, the higher absolute value of the Seebeck coefficient around the Fermi level moves from the $0\%$ strain for positive energies with a value $\vert S_{max} \vert =1.44$ mV/K and $\vert S_{max} \vert =1.00$ mV/K for negative energies. For  $+5\%$ strain in zig-zag direction we find maximum value of $\vert S_{max} \vert =0.70$ mV/K for energies below the Fermi level and above Fermi level we find a maximum $\vert S_{max} \vert =0.60$ mV/K. 
Comparing the information of the projected band structure (fig. \ref{Fig:DY}) and thermoelectric coefficients (fig. \ref{Fig:termo_DY}), we can associate that variation in the absolute value of the maximum Seebeck coefficient to a layer-selectivity in the separation of charges. Depending on the applied strain and electronic doping, the carriers tend to be mainly located in one particular layer or the other. 
Therefore, the maximum Seebeck coefficient around the Fermi level is highly determined by the properties/performance of one of the layers.

\subsubsection{Band gap selectivity: strain in armchair-direction}
Unlike the strain in the zigzag direction where we can induce selective metal-semimetal transition, 
the strain applied in the armchair x-direction locally controls the band gap. We observe how the gap 
closes or opens in some areas of the reciprocal space depending on the applied strain.
%
%Considering now the strain applied in the armchair x-direction, the band structure in fig. \ref{Fig:bnds_DX} and the thermoelectric properties in fig. \ref{Fig:termo_DX} reveal a different scenario to the zigzag case. 
%

The analysis of the projected band structure in fig. \ref{Fig:bnds_DX} reveals an exciting 
property with possible optoelectronic applications emerging from tensile strain applied in 
the x-direction (armchair direction). We can change the reciprocal space region where the band gap closes.
The sequence shown in fig. \ref{Fig:bnds_DX} shows how the metallization process can be modulated continuously upon strain, where the bands associated with the BP-layer that close the gap in the $\Gamma-X$ region for $-5$ \% armchair strain decrease in energy as the strain increases. At the same time, the bands in the $\Gamma-Y$ region separated by an energy gap moves up in energy as the strain increases and complete the gap closing for $+5$ \% strain.
This selectivity allows optical applications, using the strain-induced along in the armchair direction and light with linear polarization\cite{li2020strain,aslan2018probing}.

\begin{figure}[!h]
\centering
\includegraphics[clip,width=1.0\columnwidth,angle=0]{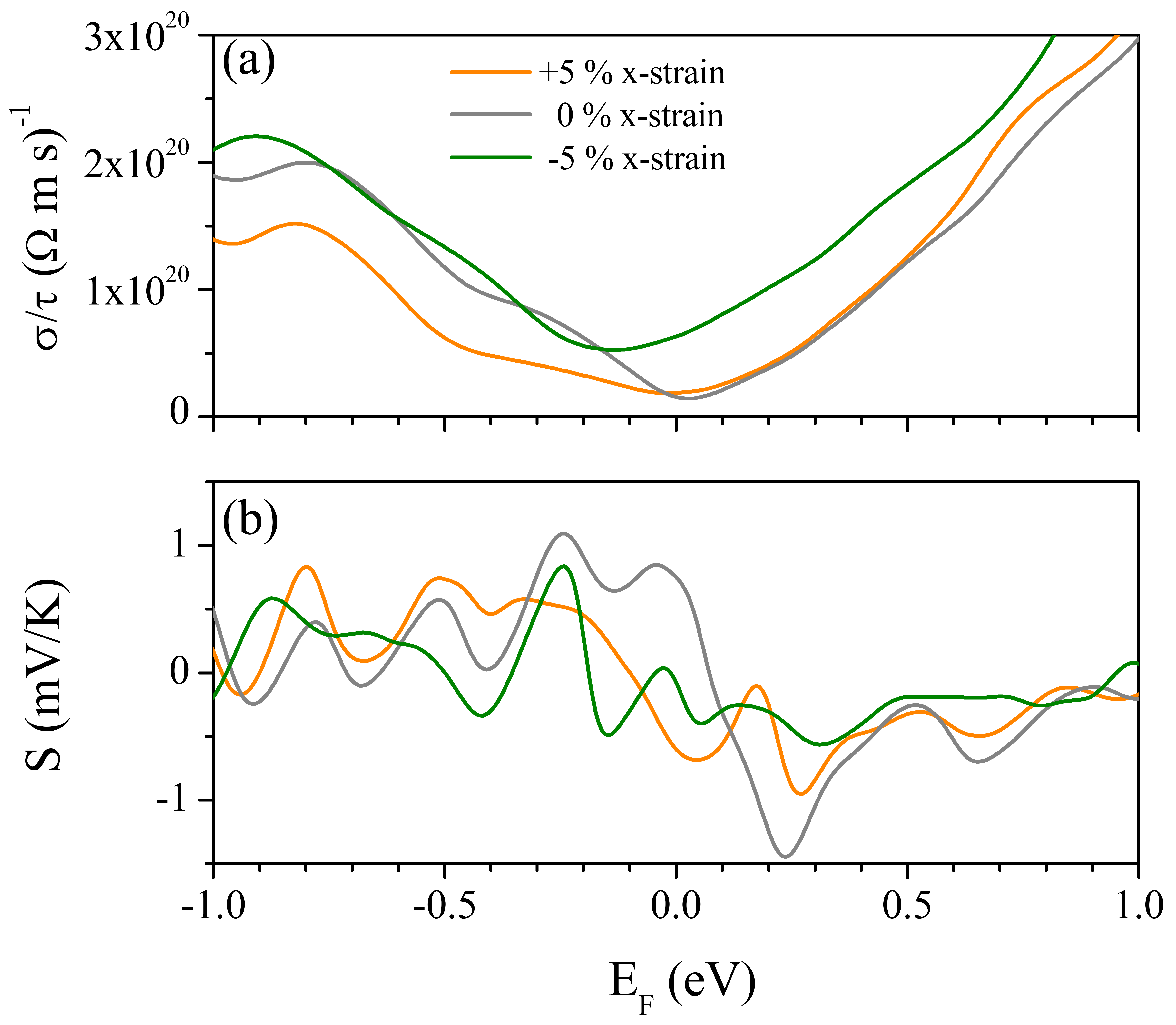}
\caption{(Color online) Thermoelectric properties for a BP-CS bilayer 
under strain in armchair direction (x-axis). 
The colors and distribution are similar to those of fig. \ref{Fig:termo_DY}, but the scale in panel (b) was adjusted. }
\label{Fig:termo_DX}
\end{figure}

The response of the thermoelectric coefficients to the applied strain in 
armchair x-direction is presented in fig.  \ref{Fig:termo_DX}.
Comparing the behavior of the electronic conductivity ($\sigma/\tau$) around the Fermi level for the different strain values (fig.  \ref{Fig:termo_DX} (a)), we notice that the curve for the $-5$ \% case presents the highest electron conductivity with a minimum of $2.6 \times 10^{19} \, \left(\Omega m s\right)^{-1}$, the minimal conductivity values for the $0$ \% and $+5$ \% cases are $0.7  \times 10^{19} \, \left(\Omega m s\right)^{-1}$ and  $0.9  \times 10^{19} \, \left(\Omega m s\right)^{-1}$ respectively.
The $3.5$ times higher electronic conductivity observed for the $-5$ \% armchair strain is associated with the bands crossing the Fermi level in the $\Gamma-X$ region. The bands belonging to the BP-layer going up from the valence band and cross with bands from the CS-layer going down from the conduction band.
For the charge neutrality point ($E=0$ eV), the conductivity for the non-strained ($0$ \%) and $+5.0$ \% cases are almost the same because the band structure does not change drastically, and the dominant features are the same.
Finally, by manipulating the electron- and hole-dopping, it is possible to adjust the conductivity of the systems under strain to follow the electronic conductivity of the non-strained case. 
For negative energies, the conductivity of the compressed systems (negative strain) approaches the non-strained curve. For positive energies, the conductivity curve of the positive strain cases follows the non-strained curve.
Our results suggest that the strain in the x-direction can be used to 
manipulate the conductivity in a predictable way. Increasing the strain 
decreases the electronic conductivity around the Fermi level.

Applying stain in the armchair direction (x-axis) decreases the extreme values of the Seebeck coefficient (fig. \ref{Fig:termo_DX} (b)) ; the curve of the non-strained system is dominant, and the cases with $\pm 5$ \% strain show a worsening of the thermoelectric performance.
Around the Fermi level, the maximum absolute value of the Seebeck coefficient corresponds to the case without strain showing a $\vert S_{max} \vert = 1.1$ mV/K for at $ E = -0.25$ eV, followed by a $\vert S_{max} \vert = 0.9$ mV/K in $0.26$ eV for strain +5\%  and finally we find that for -5\%  we have a $\vert S_{max} \vert = 0.8$ mV/K at $-0.24$ eV.
The Seebeck coefficient does not vary much when strain is applied in the armchair direction (x-axis); this phenomenon can be linked to the electronic response to the strain applied in this direction. The applied strain in the x-direction closes the band gap in one direction of the reciprocal space, but the band gap remains open in the other; for example, the gap closes in the $\Gamma-X$ region for -5\% strain, but it remains in the $\Gamma-Y$ region.

\section{Final Remarks} 
We have studied the thermoelectric properties of a Van der Waals structure composed of two different layered materials adopting a DFT approach. Both materials have a similar crystalline structure but with different strain responses;
a monolayer of black phosphorus (phosphorene) and a monolayer of carbon monosulfide.
We find a highly anisotropic response to monoaxial strain; depending on the direction, we can separately manipulate the system's electrical and thermoelectric properties. 
The selectiveness and robust response induced by the applied
strain could be used to design an active electronic component 
that works as a mechanically operated transistor with response to pressure/tension/compression.
Although the particular composition of the bilayer studied is 
not yet synthesized, our results can be interpreted as proof 
of the concept of strain-controlled layer-selective thermoelectric 
transport. 
We expect the results of this work to be easily extrapolated to other 2D materials, taking advantage of the diverse response of the band structure to uniaxial deformation.

%\section*{Acknowledgments}

The author acknowledge the financial support from ANID-FONDECYT: 
Iniciaci\'on en Investigaci\'on 2019 grant N. 11190934 (Chile) 
and USM complementary research grant year 2020 and 2021.
I thank Dr. P. Quitral-Manosalva for her critical reading of the manuscript.

%\section*{Declaration of interests}
%The authors declare that they have no known competing financial interests or personal relationships that could have appeared to influence the work reported in this paper.

%\bibliography{suya}

\end{document}